\documentstyle[12pt]{article}

\newcommand \beq{\begin{eqnarray}}
\newcommand \eeq{\end{eqnarray}}
\begin{document}


\def\nbfepsilon{\mbox{\boldmath$\epsilon$}}
\def\nbfgrad{\mbox{\boldmath$\grad$}}
\def\bfgamma{\mbox{\boldmath$\gamma$}}
\def\bfcalA{\mbox{\boldmath${\cal A}$}}
\def\bfcalS{\mbox{\boldmath${\cal S}$}}
\def\bfp{\mbox{\boldmath$p$}}
\def\bfv{\mbox{\boldmath$v$}}
\def\bfj{\mbox{\boldmath$j$}}
\def\bfhp{\mbox{\boldmath$\hat p$}}
\def\bfei{\mbox{\boldmath$e_i$}}
\def\bfe{\mbox{\boldmath$e$}}
\def\bfej{\mbox{\boldmath$e_j$}}
\def\bfk{\mbox{\boldmath$k$}}
\def\bfq{\mbox{\boldmath$q$}}
\def\bfR{\mbox{\boldmath$R$}}
\def\bfC{\mbox{\boldmath$C$}}
\def\bfR{\mbox{\boldmath$R$}}
\def\bfX{\mbox{\boldmath$X$}}
\def\bfx{\mbox{\boldmath$x$}}
\def\bfE{\mbox{\boldmath$E$}}
\def\bfB{\mbox{\boldmath$B$}}
\def\bfy{\mbox{\boldmath$y$}}
\def\bfr{\mbox{\boldmath$r$}}
\def\rmRe{\mbox{\rm$Re$}}
\def\rmIm{\mbox{\rm$ImR$}}

\def\bfgamma{\mbox{\boldmath$\gamma$}}
\def\bfalpha{\mbox{\boldmath$\alpha$}}
\def\bfsigma{\mbox{\boldmath$\sigma$}}
\def\bfalpha{\mbox{\boldmath$\alpha$}}
\def\bfsigma{\mbox{\boldmath$\sigma$}}
\def\bfSigma{\mbox{\boldmath$\Sigma$}}
\def\bfepsilon{\mbox{\boldmath$\epsilon$}}

\def\T{\hbox{temperature}}
\def\bk{\hbox{background}}
\def\P{\mbox{\psi}}
\def\BP{\mbox{\bar\Psi}}
\def\E{\hbox{equation}}
\def\Es{\hbox{equations}}
\def\QGP{\hbox{quark-gluon plasma}}
\def\HTL{\hbox{hard thermal loops}}
\def\htl{\hbox{hard thermal loop}}
\def\se{\hbox{self-energy}}
\def\pt{\hbox{polarization tensor}}
\def\pov{\hbox{point of view}}
\def\pth{\hbox{perturbation theory}}
\def\wr{\hbox{with respect to}}
\def\fn{\hbox{function}}
\def\FN{\hbox{functions}}
\def\BN{\hbox{Bloch-Nordsieck}}
\def\vp{\mbox{$\bf v\cdot p$}}
\def\vq{\mbox{$\bf v\cdot q$}}
\def\vpq{\mbox{$\bf v\cdot(p+ q)$}}
\def\tilA{\mbox{$v\cdot A$}}
\def\tilQ{\mbox{v\cdot q}}
\def\tilQ1{\mbox{$v\cdot q_1$}}
\def\tilQ2{\mbox{$v\cdot q_2$}}
\def\bfp{\mbox{\boldmath$p$}}

\def\square{\hbox{{$\sqcup$}\llap{$\sqcap$}}}   
\def\grad{\nabla}                               
\def\del{\partial}                              

\def\frac#1#2{{#1 \over #2}}
\def\smallfrac#1#2{{\scriptstyle {#1 \over #2}}}
\def\half{\ifinner {\scriptstyle {1 \over 2}}
   \else {1 \over 2} \fi}

\def\bra#1{\langle#1\vert}              
\def\ket#1{\vert#1\rangle}              

\def\simge{\mathrel{%
   \rlap{\raise 0.511ex \hbox{$>$}}{\lower 0.511ex \hbox{$\sim$}}}}
\def\simle{\mathrel{
   \rlap{\raise 0.511ex \hbox{$<$}}{\lower 0.511ex \hbox{$\sim$}}}}


\def\parenbar#1{{\null\!                        
   \mathop#1\limits^{\hbox{\fiverm (--)}}       
   \!\null}}                                    
\def\nunubar{\parenbar{\nu}}
\def\ppbar{\parenbar{p}}


\def\buildchar#1#2#3{{\null\!                   
   \mathop#1\limits^{#2}_{#3}                   
   \!\null}}                                    
\def\overcirc#1{\buildchar{#1}{\circ}{}}


\def\slashchar#1{\setbox0=\hbox{$#1$}           
   \dimen0=\wd0                                 
   \setbox1=\hbox{/} \dimen1=\wd1               
   \ifdim\dimen0>\dimen1                        
      \rlap{\hbox to \dimen0{\hfil/\hfil}}      
      #1                                        
   \else                                        
      \rlap{\hbox to \dimen1{\hfil$#1$\hfil}}   
      /                                         
   \fi}                                         %


\def\subrightarrow#1{
  \setbox0=\hbox{
    $\displaystyle\mathop{}
    \limits_{#1}$}
  \dimen0=\wd0
  \advance \dimen0 by .5em
  \mathrel{
    \mathop{\hbox to \dimen0{\rightarrowfill}}
       \limits_{#1}}}                           

\def\real{\mathop{\rm Re}\nolimits}     
\def\imag{\mathop{\rm Im}\nolimits}     

\def\tr{\mathop{\rm tr}\nolimits}       
\def\Tr{\mathop{\rm Tr}\nolimits}       
\def\Det{\mathop{\rm Det}\nolimits}     

\def\mod{\mathop{\rm mod}\nolimits}     
\def\wrt{\mathop{\rm wrt}\nolimits}     


\def\TeV{{\rm TeV}}                     
\def\GeV{{\rm GeV}}                     
\def\MeV{{\rm MeV}}                     
\def\KeV{{\rm KeV}}                     
\def\eV{{\rm eV}}                       

\def\mb{{\rm mb}}                       
\def\mub{\hbox{$\mu$b}}                 
\def\nb{{\rm nb}}                       
\def\pb{{\rm pb}}                       

%
\def\journal#1#2#3#4{\ {#1}{\bf #2} ({#3})\  {#4}}

\def\AdvPhys{\journal{Adv.\ Phys.}}
\def\AnnPhys{\journal{Ann.\ Phys.}}
\def\EurophysLett{\journal{Europhys.\ Lett.}}
\def\JApplPhys{\journal{J.\ Appl.\ Phys.}}
\def\JMathPhys{\journal{J.\ Math.\ Phys.}}
\def\LettNuovoCimento{\journal{Lett.\ Nuovo Cimento}}
\def\Nature{\journal{Nature}}
\def\NPA{\journal{Nucl.\ Phys.\ {\bf A}}}
\def\NPB{\journal{Nucl.\ Phys.\ {\bf B}}}
\def\NuovoCimento{\journal{Nuovo Cimento}}
\def\Physica{\journal{Physica}}
\def\PLA{\journal{Phys.\ Lett.\ {\bf A}}}
\def\PLB{\journal{Phys.\ Lett.\ {\bf B}}}
\def\PR{\journal{Phys.\ Rev.}}
\def\PRC{\journal{Phys.\ Rev.\ {\bf C}}}
\def\PRD{\journal{Phys.\ Rev.\ {\bf D}}}
\def\PRB{\journal{Phys.\ Rev.\ {\bf B}}}
\def\PRL{\journal{Phys.\ Rev.\ Lett.}}
\def\PhysRept{\journal{Phys.\ Repts.}}
\def\ProcNatlAcadSci{\journal{Proc.\ Natl.\ Acad.\ Sci.}}
\def\ProcRoySoc{\journal{Proc.\ Roy.\ Soc.\ London Ser.\ A}}
\def\RevModPhys{\journal{Rev.\ Mod.\ Phys. }}
\def\Science{\journal{Science}}
\def\SovPhysJETP{\journal{Sov.\ Phys.\ JETP }}
\def\SovPhysJETPLett{\journal{Sov.\ Phys.\ JETP Lett. }}
\def\SovJNuclPhys{\journal{Sov.\ J.\ Nucl.\ Phys. }}
\def\SovPhysDoklady{\journal{Sov.\ Phys.\ Doklady}}
\def\ZPhys{\journal{Z.\ Phys. }}
\def\ZPhysA{\journal{Z.\ Phys.\ A}}
\def\ZPhysB{\journal{Z.\ Phys.\ B}}
\def\ZPhysC{\journal{Z.\ Phys.\ C}}

\begin{titlepage}
\begin{flushright} {Saclay-T95/146}
\end{flushright}

\vspace*{0.2cm}
\begin{center}
\baselineskip=20pt {\Large LIFETIME OF QUASIPARTICLES  IN HOT QED PLASMAS}

 \vskip0.5cm Jean-Paul
BLAIZOT\footnote{CNRS}  and Edmond IANCU\footnote{CNRS}\\
 {\it Service
de Physique Th\'eorique\footnote{Laboratoire de la Direction des Sciences
de
la Mati\`ere du Commissariat \`a l'Energie Atomique}, CE-Saclay \\ 91191
Gif-sur-Yvette, France}

\end{center}

\vskip 1cm
\begin{abstract}
The calculation of the lifetime of quasiparticles in a QED plasma at high
temperature remains plagued with infrared divergences, even after one has taken into
account the screening corrections. The physical processes 
responsible for these
divergences are the collisions involving the exchange of
very soft, unscreened, magnetic photons, whose contribution 
 is enhanced by the thermal
Bose-Eisntein occupation factor. The self energy diagrams which
diverge in perturbation theory contain  no internal fermion
loops, but an arbitrary number of internal magnetostatic photon
lines. By generalizing the Bloch-Nordsieck model at finite
temperature, we can resum all the singular contributions of such
diagrams, and obtain  the correct long time behaviour of the
retarded fermion propagator in the hot QED plasma:
$S_R(t)\sim
\exp\{-\alpha T \, t\, \ln\omega_pt\}$, where $\omega_p=eT/3$ is the plasma frequency
and $\alpha=e^2/4\pi$.
\end{abstract}

\vskip 1.6cm

\end{titlepage}

\baselineskip=25pt
\setcounter{equation}{0}

It is commonly assumed that the notion of quasiparticles is a useful concept in the
description of ultrarelativistic plasmas. This implies that the damping rate
$\gamma$  of quasiparticle excitations, obtained from the exponential decay of the
 retarded propagator, 
 $S_R(t,{\bf p})\,\sim\,{\rm e}^{-i\omega({\bf p}) t} {\rm e}^{ -\gamma({\bf p}) t}$, 
is small compared to the quasiparticle energy $\omega({\bf p})$. In hot gauge
theories, the typical energy of the relevant
quasiparticles is the temperature  $T$, while one expects
$\gamma\sim g^2T$ \cite{Pisarski89,BP90,Lebedev90}, where
$g$ is the gauge coupling (in QED, $g=e$ is the electric charge).
The same damping rate is expected for the collective excitations, 
whose typical energies are $\sim gT$ \cite{BP90}. This suggests that, indeed, in
the weak coupling regime, quasiparticles are well defined, and
collective modes are weakly damped. However, the computation of 
$\gamma$ in perturbation theory 
is plagued with infrared divergences, which casts doubt on the
validity of these statements [1-7]. We shall show in this paper that the
infrared divergences which remain after the screening corrections
have been taken into account are due to collisions involving the
exchange of very soft, unscreened, magnetic photons. Such
divergences occur in all orders of perturbation theory and can
 be eliminated only by a non perturbative calculation of
the quasiparticle  propagator.  We present in this letter such a non
perturbative analysis, based on an extension of the Bloch-Nordsieck
model at finite temperature. The final result is that
quasiparticles do indeed exist, although they do not correspond to
the usual exponential decay indicated above, but to a more
complicated behaviour,
$S_R(t,{\bf p})\,\sim\,{\rm e}^{-i\omega({\bf p}) t} {\rm e}^{-\alpha T \, t\,
\ln\omega_pt}$, where $\omega_p$ is the plasma frequency. 

We shall consider in most of this letter the case of hard 
fermions, with momenta $p\sim T$, and restrict ourselves to QED. (The case of
collective excitations will be discussed briefly at the end.) In perturbation theory,
 the damping rate is obtained from the imaginary part
of the self-energy on the unperturbed mass-shell $\omega=p$ :
\beq\label{damping}
\gamma\,=\, -\frac{1}{4p}\,{\rm tr}\left(
{\slashchar p}\, {\rm Im}\,\Sigma(\omega+i\eta,{\bf p})\right)\Big |_
{\omega=p}\,.\eeq 
To bare one-loop order, $\gamma=0$ by kinematics : the on-shell fermion
 can emit (or absorb) only a space-like
photon ($q^2\equiv q_0^2 - {\bf q}^2 \,\le 0$), and there are no such photons at
tree level (a more complete discussion of this point will be presented in a longer
publication \cite{next}). However, through the Landau damping mechanism, the spectral
function of a photon in the plasma does acquire a non vanishing  weight for
 space-like momenta \cite{Weldon82}. One takes this into account by resumming the
so-called hard thermal loops (HTL)  in the internal
photon line of the one loop self-energy, and obtains in this way the leading
contribution to the damping rate \cite{Pisarski89,BP90}. Physically, this
contribution represents  the effect of the collisions induced by one photon
exchange \cite{Baym90}. Note that the corresponding collision rate would be 
 quadratically divergent without the screening corrections
 \cite{Baym90,BThoma91}. Such
corrections are sufficient to
render finite the transport 
coefficients \cite{Baym90}, or the collisional energy 
loss \cite{BThoma91}, but not the total interaction rate \cite{Heisel92}, nor the
damping rate [1-7]. These quantities, except for the damping rate
of a quasiparticle with zero momentum \cite{BP90,Kobes92},  remain
logarithmically divergent,  both in  abelian and non-abelian gauge
theories. In QCD, this problem is commonly bypassed 
 by advocating the  IR cut-off provided by a possible 
magnetic  mass $\sim g^2T$ (see e.g. \cite{Pisarski93}). But such a solution cannot
apply for 
 QED where one expects  no magnetic
screening \cite{Fradkin65}. 

In the imaginary time formalism, and in the Coulomb gauge
(the one-loop result for $\gamma$ is gauge 
independent\cite{Rebhan93}), the resummed one loop 
 diagram is evaluated as \cite{Pisarski93}
\beq\label{Sigeff}
{}^*\Sigma(p)&=&-\,g^2 T\sum_{q_0= i\omega_m}
 \int \frac{{\rm d}^3q}{(2\pi)^3}\,
\,\gamma_\mu\,S_0(p+q)\,\gamma_\nu
\,{}^*D^{\mu\nu}(q)\,\nonumber\\
&=& -\,g^2 \int \frac{{\rm d}^3q}{(2\pi)^3}
\int_{-\infty}^{+\infty}\frac{dk_0}{2\pi}
\int_{-\infty}^{+\infty}\frac{dq_0}{2\pi}
\rho_0(k)\gamma_\mu {\slashchar k} \gamma_\nu\,{}^*\rho^{\mu\nu}(q)\,
\frac{N(q_0)+n(k_0)}{k_0-q_0-p_0}\,.\eeq
In this equation,   ${\bf k=p+q}$, $p_0= i\omega_n= i(2n+1)\pi T$,
and  $\omega_m= 2\pi mT$, with integers $n$ and $m$. We denote by 
 $\rho_0(k_0, k)$ the spectral density of the (hard) fermion,
and by ${}^*\rho_{l,t}(q_0,q)$ 
those of the (soft) longitudinal ($l$) and transverse ($t$) photons.
According to eq.~(\ref{damping}), one obtains the damping rate $\gamma$ 
 from the imaginary part of ${}^*\Sigma$, 
after analytical continuation of $p_0$ to the real axis $(p_0\to
\omega+i\eta)$. One thus gets $\gamma(\omega \simeq
p)=g^2p\,\Phi(\omega, p)$, where ($\cos \theta= \hat{\bf p}\cdot
\hat {\bf q}$)
\beq\label{phi}
\Phi(\omega, p)&\equiv&
 \int \frac{{\rm d}^4q}{(2\pi)^4}  \int \frac{{\rm d}^4k}{(2\pi)^4}\,
(2\pi)^4\delta^{(4)}(p+q-k)\,
\nonumber \\ &&\Bigl[N(q_0)+n(k_0)\Bigr]
\rho_0(k)\,
\left({}^*\rho_l(q)+(1-\cos^2\theta)
{}^*\rho_t(q)\right)\,,\eeq
is the thermal phase-space  for the decay of the fermion.
The infrared singular region is that of small photon
energy-momentum $q_0,\,q\ll gT$. Since  
 $\omega, \,p \sim T$, in studying this region, one  can expand
$|{\bf p+q}|\simeq p+q\cos \theta$ and
$N(q_0)+n(k_0)\simeq N(q_0) \simeq T/q_0 \gg 1$. One obtains then
\beq\label{gammap}
\Phi(\omega, p)&=& \frac{T}{2p}
\int \frac{{\rm d}^3q}{(2\pi)^3}
 \int_{-\infty}^\infty \frac{{\rm d}q_0}{2\pi q_0}
2\pi \delta(\omega-p +q_0 -q\cos \theta)
\left({}^*\rho_l(q)+(1-\cos^2\theta)
{}^*\rho_t(q)\right)\nonumber\\
&=& \frac{T}{4\pi p}
\int_{0}^{\infty}{\rm d}q \,q\,\theta (q-\delta E)
 \int_{-q}^q\frac{{\rm d}q_0}{2\pi q_0}
\left\{{}^*\rho_l(q_0,q)\,+\,\left(1-\frac{q_0^2}{q^2}\right)
{}^*\rho_t(q_0,q)\right\}\,,\eeq
where the   $\delta$-function 
has been used to perform the angular integration,
and  $\delta E\equiv |\omega-p| >0$.
Note the restriction to space-like momenta,  $|q_0|\le q$:
only the off-shell  photons contribute to the fermion damping, 
which is consistent with the physical interpretation of
$\gamma$ in terms of collisional processes, as alluded to earlier.
The evaluation of the $q_0$-integral can be done easily by using
sum rules satisfied by the spectral function \cite{Pisarski93}.
The singular contribution comes from the transverse component, for
which we have
\beq\label{SumR}
\int_{-q}^q\frac{{\rm d}q_0}{2\pi q_0}\,
{}^*\rho_t(q_0,q)=\frac{1}{q^2}\,-\,\frac{z_t(q)}{\omega_t^2(q)},\eeq
where ${\omega_t(q)}$ and $ z_t(q)$ are respectively
the position and the residue of the transverse plasmon pole.
As $q\to 0$,  ${z_t(q)}/{\omega_t^2(q)} \to 1/\omega_p^2 $, so that
for $q\ll \omega_p\sim gT$
the integral is dominated by the term $1/q^2$.
In fact, the function ${}^*\rho_t(q_0,q)/q_0$ 
is strongly peaked at $q_0=0$, and in the calculation
of the integral (\ref{SumR}) it can be replaced by the following
approximate expression:
\beq\label{rhot}
\frac{1}{q_0}\,{}^*\rho_t(q_0\ll q)= \frac{3\pi}{2}\, \frac{\omega_p^2 \, q}
{q^6\,+\, (3\pi \omega_p^2 q_0/4)^2}\,\longrightarrow_{q\to 0}
\,\frac{2\pi}{q^2}\,\delta(q_0).\eeq
Keeping only this leading term, and ignoring the non-singular
contribution of the electric modes, one finds
\beq\label{gammap1}
\gamma (\delta E)=g^2 p\,\Phi(\delta E)\simeq \frac{g^2T}{4\pi}
\int_{0}^{\omega_p}{\rm d}q \,q\theta (q-\delta E)\,\frac{1}{q^2}
= \alpha T \ln \frac{\omega_p}{\delta E},\eeq
with $\alpha = g^2/4\pi$. Thus, the singularity of $\gamma$ when
$\delta E=0$, i.e., when 
the fermion is on-shell, can be attributed 
to the  low frequency magnetic photons produced by Landau damping, whose
contribution is enhanced by the Bose-Einstein occupation factor $T/q_0$. Note that
since the energy conservation requires
$q_0=q\cos\theta$, these photons are 
emitted, or absorbed,  at nearly 90 degrees. 

The upper cut-off $\omega_p\sim gT$  in eq.~(\ref{gammap1}) 
 accounts approximately for the terms which have been
neglected when keeping only the $1/q^2$ contribution to the
 r.h.s of eq.~(\ref{SumR}): as $q\gg \omega_p$,
 the r.h.s. of eq.~(\ref{SumR})
is decreasing like $\omega_p^2/q^4$, so that the exact integrand
in eq.~(\ref{gammap}) is indeed cut-off at $q\sim \omega_p$. 
Note that, as long as we are interested only in
the coefficient of the logarithm,
 the precise value of this cut-off is unimportant. The scale $\omega_p$ however
is uniquely determined by the physical process responsible for the existence
of space like photons, i.e., the Landau damping. This is important since, as we
shall see later, it is the scale which fixes the long time behavior of the retarded
propagator.

In the forthcoming discussion, we shall take advantage of a major technical
simplification which is based on the following observation: the
logarithm in eq.~(\ref{gammap1}) arises entirely from the magnetic contribution of
the term
$q_0=i\omega_m=0$  in the Matsubara sum  of eq.~(\ref{Sigeff})
\cite{Burgess92,Rebhan95}. The analytic continuation of this term to real energy (
$p_0\to \omega+i\eta$) is well defined, and it yields the
following  imaginary part 
\beq\label{gammaIR}
- \frac{1}{4p}\,{\rm tr}\left(
{\slashchar p}\,{\rm Im}\, {}^*\Sigma_0(p)\right)&\simeq&
-2 g^2 \omega T\int \frac{{\rm d}^3q}{(2\pi)^3}\,\frac{1}{q^2}
\, {\rm Im}\,\frac
{1}{(\omega+i\eta)^2-({\bf p+q})^2}\nonumber\\&\simeq&
 \frac{g^2T}{4\pi}
\int_{0}^{\omega_p}\frac{{\rm d}q}{q}\theta (q-\delta E),\eeq
where the approximate equality means that only regular terms have 
been dropped. As anticipated, this coincides with eq.~(\ref{gammap1}).
Note that, since the static transverse propagator
is not modified by the HTL corrections,  eq.~(\ref{gammaIR}) 
could have been obtained directly from the term  
 $q_0=i\omega_m=0$ in the {\it bare} one-loop self-energy,
provided one introduces the same upper cut-off $\omega_p$ in the
$q$-integral.

Some remarks are needed at this stage.
 At the bare one-loop level, $\gamma =0$, as mentioned.
Thus, in the bare one loop calculation, there is a cancellation
between the logarithmic divergence of the static mode, and that of
the {\it sum of all} the non-static ones \cite{next}. The inclusion of the
screening corrections on the photon line removes the divergence of
the contribution of the nonstatic modes, leaving a divergence 
in the static contribution only. When $\omega\to p$, the integrand in
eq.~(\ref{gammaIR}) becomes proportional to $\delta(\cos\theta)$, so that the
singular contribution of the static mode originates from the emission or absorption
of static magnetic photons at 90 degrees. As we have seen, these processes represent
the dominant contribution of the resummed one loop calculation. That they can be taken
into account by what looks like a bare one loop calculation is due to the fact that
the  integrated transverse photon spectral weight  (\ref{SumR})  is
$1/q^2$ at small
$q$, i.e., is indistinguishable from the propagator of a static
 photon. Finally, as already mentioned, the source of the
space-like photons is Landau damping; this is the physical origin of
the  cut-off $\omega_p$ to be imposed on the one loop integral of
eq.~(\ref{gammaIR}).

Power counting indicates that mass-shell divergences
do occur in higher orders as well. The most singular diagrams are those where
the fermion propagator is dressed by static magnetic
photon lines. When evaluated on the tree-level mass-shell
$\omega=p$, and in the presence of an IR regulator $\mu$, such
diagrams generate contributions of the type
$(g^2T/\mu)^{n-1}$, where $n$ is the number of loops.
 Such power-like IR divergences are analogous to those identified in the
analysis of the corrections to the screening mass in \cite{debye}, and their
presence signals a breakdown of perturbation theory. The calculation of the
fermion propagator in the IR
singular domain requires therefore a non perturbative approach, to
which we now turn. 

Since internal fermion loops are not essential
anymore at this stage, we may restrict ourselves to the ``quenched
approximation'' involving only magnetostatic photons
\cite{debye}. In this approximation, the imaginary-time
 fermion propagator can be written as the following
functional integral
\beq\label{FS}
S(x,y)= Z_0^{-1}\int [{\rm d} {\bf A}]\, G(x,y|{\bf A})\,{\rm
exp}\left\{-
\frac{1}{2}\,\Bigl({\bf A}, D_0^{-1}{\bf A}\Bigr)_0\right\},\eeq
 where  $ G(x,y|{\bf A})$ is the fermion propagator in the presence
of a background gauge field, and 
\beq\label{SEFF} 
\Bigl({\bf A},D_0^{-1}{\bf A}\Bigr)_0\,=\,\frac{1}{T} \int {\rm
d}^3x\, {\rm d}^3y\,A^i({\bf x})\,
D^{-1}_{0\,ij} ({\bf x}-{\bf y}) A^j ({\bf y}).\eeq
In this equation, the factor $1/T$ has its origin in the restriction to the
zero Matsubara frequency. The propagator  
$D_{0\,ij}$ is that of a free static photon,
 \beq\label{D0}
D^{ij}_0({\bf x})\equiv
\int \frac{{\rm d}^3q}{(2\pi)^3}\,
{\rm e}^{i{\bf q\cdot x}}\, D^{ij}_0({\bf q})
\eeq
with
\beq
D_0^{ij}({\bf q})
= \,\frac{\delta^{ij}}{q^2}+(\lambda -1)\frac{q^iq^j}{q^4},
 \eeq
in an arbitrary Coulomb or covariant gauge ($\lambda=0$
corresponds to both the Landau and the
 strict Coulomb gauges).

The fermion propagator in a {\it static} background  field depends only on the time
difference $x_0-y_0$,
$G(x,y|{\bf A})\equiv G(x_0-y_0, {\bf x, y}|{\bf A})$.
Its Fourier transform can be analytically continued
in the complex energy plane, and the resulting function coincides,
in the upper half plane, with the retarded propagator.
It is then convenient to take the Fourier transform of
eq.~(\ref{FS}).
To this aim, we write ($p^\mu=(i\omega_n,{\bf p})$, $\omega_n=(2n+1)\pi T$):
\beq\label{ansatz}
G(x,y|{\bf A})=T\sum_{n} \int \frac{{\rm d}^3p}{(2\pi)^3}\,
{\rm e}^{-ip\cdot(x-y)}G(p;{\bf x}|{\bf A}),
\eeq
and similarly for $S$. We obtain
\beq\label{SEP} S(p)
\equiv Z_0^{-1}\int [{\rm d} {\bf A}]
\,G(p;{\bf x}|{\bf A}) \,{\rm exp}\left\{-
\frac{1}{2}\,
\Bigl({\bf A}, D_0^{-1}{\bf A}\Bigr)_0\right\}.\eeq
 Since the energy 
$p_0$ enters eq.~(\ref{SEP}) as an external parameter, the
continuation to real external energy $p_0\to \omega+i\eta$, and the
Fourier transform
 to real time, 
can both be performed {\it before} doing the functional integration \cite{next}. 
This offers the possibility to calculate the retarded
propagator $S_R(t,{\bf x})$ by directly inserting an approximate expression for 
$G_R(x,y|A)$ in the functional integral (\ref{FS}).

In the kinematical regime of interest, an approximate expression
for 
$G_R(x,y|{\bf A})$ is obtained by neglecting the recoil of the
fermion in the successive emissions or absorptions of very
soft photons.  More precisely, we note that inside the diagrams generated by
eq.~(\ref{FS}), we can approximate  the fermion
propagators by
\beq \label{SIR}
S_0(\omega, {\bf p+q})\,=\,
\frac{-\omega  \gamma_0 + ({\bf p+q})\cdot \bfgamma}
{(\omega+i\eta)^2-\epsilon_{p+q}^2}\,\simeq\,
\frac{-1}{\omega -\epsilon_p - {\bf v}\cdot {\bf q} +i\eta}\,
\frac{\gamma_0- \hat{\bf p}\cdot \bfgamma}{2},
\eeq
where ${\bf q}$ is a linear combination of the internal photons
momenta and ${\bf v}\equiv \del\epsilon_p/\del {\bf p}$
(${\bf v}=\hat{\bf p}$ for the ultrarelativistic fermion).
This is the familiar structure encountered in most treatments
of IR divergences in QED (see, e.g., \cite{Weinberg}) and which is
 economically exploited within  the Bloch-Nordsieck model
(see, e.g. \cite{Bogoliubov}). In this model, the propagator
with the desired IR structure  is obtained in
 coordinate space as the solution to
\beq\label{freeprop}
-i\,(v\cdot \del_x)\,G_0(x,y)&=&\delta(x-y),\eeq
where $v^\mu  =(1,{\bf v})$. In the presence of an arbitrary background field, the
corresponding propagator, $ G(x,y|A)$, satisfies 
the following equation ($D_\mu=\del_\mu +igA_\mu$)
\beq\label{Gret}
-i\,(v\cdot D_x)\,G(x,y|A)&=&\delta(x-y),\eeq
which  can be solved {\it exactly}. For retarded boundary conditions,
and for static fields:
\beq\label{GR}
G_{R}(x,y|{\bf A})&=&i\,\theta (x_0-y_0)\,\delta^{(3)}
\left({{\bf x}}-{{\bf y}}-{{\bf v}}(x_0-y_0)
\right )U(x,y)
\eeq
where $U(x,y)$ is the parallel transporter 
\beq\label{U} 
U(x,x-vt)=\exp\left\{ ig\,\int_0^t {\rm d}s\, {\bf v}
\cdot {\bf A}({\bf x-v}s)
\right\}.\eeq

The retarded propagator $S_R(x-y)$ is calculated by inserting 
the expressions 
(\ref{GR}), (\ref{U}) of $G_{R}(x,y|{\bf A})$ in  the functional
 integral (\ref{SEP}). It can be written as 
\beq\label{SRT}
S_R(t,{\bf p})&=&i\theta(t) {\rm e}^{-it({\vp})}\,\Delta(t),\eeq
where
\beq\label{Delta0}
\Delta(t)\equiv Z_0^{-1}\int [{\rm d} {\bf A}]
\,U(x,x-vt) \,{\rm exp}\left\{-
\frac{1}{2}\,\Bigl({\bf A}, D_0^{-1} {\bf A}\Bigr)_0\right\}\eeq
contains all the non-trivial time dependence. The functional integral is easily
done:
\beq\label{Delta1}
\Delta(t) ={\rm exp}\left\{\frac{g^2}{2}\,T\int_0^t {\rm d}s_1 
\int_0^t {\rm d}s_2\, v^i\, D^{ij}_0({\bf v}(s_2-s_2))\, v^j
\right \}.\eeq
In this equation, $D^{ij}_0({\bf x})$ is the coordinate space representation of the
magnetostatic photon propagator (see eq~(\ref{D0})). 

The $s_1$ 
and $s_2$ integrations in eq.~(\ref{Delta1}) can be done by going to the Fourier
representation:
\beq\label{SR0}
\Delta(t)=  {\rm exp}\left \{-g^2T
\int \frac{{\rm d}^3q}{(2\pi)^3} 
\frac{\tilde D({\bf q})}{({\vq})^2}\,
\Bigl(1-  {\rm cos}\,t({\vq})\Bigr)\right\},\eeq
where $ v^i\, D_0^{ij}({\bf q}) v^j\equiv \tilde D({\bf q})$. 
The integral in eq.~(\ref{SR0}) is identical to that one would 
get in the Bloch-Nordsieck model in 3 dimensions.
It has no infrared divergence, but one can verify that the expansion of  
$\Delta(t)$ in powers of $g^2$ generates 
the most singular pieces of the usual perturbative expansion
for the self-energy\cite{next}. One can also verify that the integral in
eq.~(\ref{SR0}) presents an ultraviolet logarithmic divergence. However, one should
recall that the restriction to 
 the static photon mode implies that such an integral is to be cut off
at momenta $q\sim \omega_p$ (cf.  the discussion after
eq.~(\ref{gammap1})).  

The calculation of $\Delta(t)$ is most simply done using the
coordinate space representation (\ref{Delta1}). In the Feynman
 gauge $\lambda =1$,
we use 
$ v^i\,  D^{ij}_0({\bf x})\, v^j\,=\, 1/4\pi x$ and obtain 
\beq\label{qint}
 &\, & \frac{1}{2} \,\int_0^t {\rm d}s_1 
\int_0^t {\rm d}s_2\, v^i\, D_0^{ij}({\bf v}(s_2-s_2))\, v^j
\nonumber\\ &\, & =
\frac{1}{8\pi}\int_0^t {\rm d}s_1 
\int_0^t {\rm d}s_2\, \frac{\theta(|s_1-s_2|-1/\omega_p)}{|s_1-s_2|}
\simeq\,\frac{t}{4\pi}\,\ln \omega_pt .
\eeq
where the  ultraviolet cut-off is introduced in the function
$\theta(|s_1-s_2|-1/\omega_p)$. This expression is valid for times 
$t\gg 1/\omega_p$. It is insensitive to the specific procedure which is
used to implement the cut-off \cite{next}, and independent of gauge fixing (the
gauge-dependent contribution to the integral above, as given by the last term of the
photon propagator (\ref{D0}), is $(\lambda-1)\,\frac{t}{8\pi}$, and therefore
subleading).

At times $t\gg 1/\omega_p$ the function $\Delta(t)$ is thus  of
the form \cite{Takashiba95}
\beq\label{DLT}
\Delta(\omega_pt\gg 1)\simeq {\rm exp}\Bigl( -\alpha Tt \ln \omega_p
t\Bigr).\eeq
 A measure of the decay time $\tau$ is given by 
\beq \frac{1}{\tau}=\alpha T\ln \omega_p \tau=
\alpha T\left(\ln \frac{\omega_p}{\alpha T} - \ln
\ln \frac{\omega_p}{\alpha T} + \,...\right).\eeq
Since $\alpha T \sim g\omega_p$,  $\tau \sim
 1/(g^2 T \ln (1/g))$. This corresponds to a damping rate 
$\gamma\sim1/\tau\sim g^2 T\ln (1/g)$, similar to that obtained in a
one loop calculation  with an IR cut-off
$\sim g^2T$ (cf. eq.~(\ref{gammap1})).

However, contrary to what perturbation theory predicts, 
$\Delta(t)$ is decreasing faster than any exponential. It follows that
 the  Fourier transform 
\beq\label{SRE}
S_R(\omega, {\bf p})\,=\,
\int_{-\infty}^{\infty} {\rm d}t \,{\rm e}^{-i\omega t}
S_R(t,{\bf p})\,=\,
i\int_0^{\infty}{\rm d}t
\,{\rm e}^{it(\omega- {\bf v\cdot p}+i\eta)}\,\Delta(t),\eeq
  exists 
for {\it any} complex (and finite) $\omega$. Thus,  the retarded propagator
 $S_R(\omega)$ is an entire function, with sole singularity at 
Im$\,\omega\to -\infty$. 

The previous analysis can be extended \cite{next} to quasiparticles
 with {\it soft} momenta, $p\sim gT$, for which the same infrared 
difficulty arises
\cite{Pisarski93}. Because of the collective nature
 of the soft quasiparticles, the
HTL  resummations to be performed in this case are more elaborate
\cite{BP90}.  It can be shown however that the leading divergences
are  again confined to the magnetostatic photon
sector\cite{Rebhan95}, and occur in diagrams which  involve only the
 3-point photon-fermion vertex
\cite{next} (to one-loop order, this has been noticed already in
Refs.\cite{Pisarski93,Rebhan95}). 
In the kinematical regime of interest, one finds
\cite{next} that the approximate fermion propagator and 3-point vertex are of the
form:
\beq
\label{effFeyn}
S_{\pm}(\omega,{\bf p+q})=\frac{-z_{\pm}(p)}{\omega-\omega_{\pm}(p)
-{\bf v}_{\pm}(p)\cdot {\bf q}}\,\frac{\gamma_0\mp \hat{\bf p}\cdot
\bfgamma}{2},\qquad\Gamma_{\pm}^i=
\frac{v^i_{\pm}}{z_{\pm}}\,\frac{\gamma_0\pm \hat{\bf p}\cdot
\bfgamma}{2}.\eeq
In these equations, the subscripts $\pm$ refer to the
two positive-energy modes of the soft electron\cite{Klimov81},
with energies $\omega_{\pm}(p)$ and residues $z_{\pm}(p)$, and ${\bf v}_{\pm}(p)
\equiv \del \omega_{\pm}(p)/\del {\bf p}$. The main approximation used in the
present paper  remains meaningful for a soft  quasiparticle whose 
 momentum $p\sim gT$ is much larger than the typical
momenta $q\simle g^2T$ of the off-shell photons which
are responsible for the leading IR singularity.
By the  same steps as before, we obtain the retarded propagator for the two
fermionic modes 
$\pm$ in the form
\beq\label{SRpm}
S_{\pm}(\omega, {\bf p}) &=&i\,z_{\pm}(p)\int_0^{\infty}{\rm d}t
\,{\rm e}^{it(\omega- \omega_{\pm}(p)+i\eta)}\,\Delta_{\pm}(t),
\nonumber\\ \Delta_{\pm}(t)&= &\Delta (|v_{\pm}|t),\eeq
with the function $\Delta(t)$ given by eq.~(\ref{SR0}).

To conclude, we have shown that, in the weak coupling regime, quasiparticles
excitations in the ultrarelativistic QED plasma are slowly damped. 
 Their spectral density retains the shape
of a {\it resonance}  strongly peaked around the perturbative mass-shell 
$\omega \sim {\vp}$, 
with a typical width of order $\sim g^2T \ln(1/g)$. No singularity is
associated to this resonance: the retarded propagator is an entire function in the
complex energy plane.

\end{document}